\documentclass[letter]{aa}
\usepackage{graphicx}

\usepackage{txfonts}
 \usepackage{newclude}
\usepackage[round]{natbib}
\bibpunct{(}{)}{;}{a}{}{,}
\graphicspath{{./images/}}
\begin{document}

\title{The GALEX Ultraviolet Virgo Cluster Survey (GUViCS) VIII. Diffuse dust in the Virgo intra-cluster space}

\author{A. Longobardi\inst{1}, A. Boselli\inst{1}, S. Boissier\inst{1}, S. Bianchi\inst{2}, P. Andreani\inst{3}, E. Sarpa\inst{1}, A. Nanni\inst{1}, M. Miville-Desch\^{e}nes\inst{4} }
\offprints{A. Longobardi}

\institute{Aix Marseille Univ, CNRS, CNES, LAM, Laboratoire d'Astrophysique de Marseille,  Marseille, France\\ e-mail:
  alessia.longobardi@lam.fr
\and INAF-Osservatorio Astrofisico di Arcetri, Largo E. Fermi 5,I-50125, Florence, Italy 
\and European Southern Observatory, Karl-Schwarzschild-Strasse 2, 85748 Garching, Germany
\and Laboratoire AIM, CEA / CNRS / Universit\'{e} Paris-Saclay, 91191, Gif-sur-Yvette, France}

\date{Draft......, Received .......; Accepted .......}

   \authorrunning{A.Longobardi et al.}
   \titlerunning{Diffuse dust component of the Virgo cluster}

 
\abstract 
{}
{We present the first detection of diffuse dust in the intra-cluster medium of the Virgo cluster out to $\sim$0.4 virial radii, and study the radial variation of its properties on a radial scale of the virial radius.}
{Analysing near-UV - $i$ colours for a sample of $\sim12000$ background galaxies with redshifts $0.02 < z < 0.8$, we find significant colour reddening and relate it to variation in $E(B-V)$ values. }
{The $E(B-V)$ mean profile shows a dust component characterised by an average reddening $E(B-V)\sim0.042 \pm 0.004$ mag within 1.5 degrees ($\sim0.3\, r_{vir}$) from the cluster centre. Assuming a Large Magellanic Cloud extinction law, we derive an average visual extinction $A_{V} = 0.14\pm 0.01$ for a total dust mass,  $M_{d} = 2.5\pm0.2\times10^{9}M_{\odot}$, hence a dust-to-gas mass ratio $M_{d}/M_{g} = 3.0\pm 0.3 \times 10^{-4}$. Based on the upper limits on the flux density $\mathrm{I_{250\mu m} = 0.1\, MJy sr^{-1}} $ derived from $Herschel$ data, we estimate an upper limit for the dust temperature of $T_{d} \sim 10\, K$. However, similar densities can be obtained with dust at higher temperatures with lower emissivities.}
{The Virgo cluster has diffuse dust in its intra-cluster medium characterised by different physical properties as those characterising the Milky Way dust. The diffuse dust in Virgo is transported into the cluster space through similar phenomena (stripping) as those building up the optical intra-cluster light, and it constitutes an additional cooling agent of the cluster gas.}

   \maketitle
%


\section{Introduction} 
It is now well known that a fraction of the baryonic content in galaxy clusters is represented by the intra-cluster light (ICL), a stellar population that is gravitationally bound to the cluster potential, whose production is thought to be tightly linked to the evolution of galaxies in clusters \citep[e.g.][]{dolag10,contini14}.
While the IC population is an important component to study and its existence is well established, it has historically been proved difficult to detect because its surface brightness is low, so that no uniform information has been gathered so far on its properties across the electromagnetic spectrum. IC stars are studied through their optical properties \citep[e.g.][]{gonzales05,mihos17}, and near-infrared (NIR) analyses of optically identified IC features show that these have NIR emission \citep{krick11}. Finally, the hot IC plasma, or ICM, is studied through X-ray data \citep[e.g.][]{sarazin86,mohr99,neumann05}. However, little is known about its properties in the form of warm or cold gas and dust. 
Dust in the intra-cluster space (intra-cluster dust, ICD) is expected to survive sputtering by the harsh X-ray emitting gas for a typical timescale of $\sim 10^{8}$ yr \citep{ds79}. Detected emission would then provide a direct clue to the different injection mechanisms and their efficiencies \citep[primordial infalling of the diffuse intergalactic dust component, galactic winds, mass loss from ICM red giants and supergiants;][]{popescu00,shindler05,domainko06}. Moreover, it may trace the current accretion rate of the cluster \citep{popescu00}. 
Researchers have therefore tried to detect the ICD mainly through far-IR (FIR) or submillimeter studies that directly search for thermal dust emission or through an indirect search for extinction and reddening of background sources \citep[e.g.][]{giard08,gutierrez17}, but the results are still controversial.
The diffuse ICD is found to produce flux density levels at 70 $\mu $m of $\sim 0.06-0.1\, \rm{MJy\, sr^{-1}}$ in nearby clusters such as Coma or Perseus, even though evidence of dust grains in the IC space has also been collected for systems at higher red-shifts, with flux emissions expected to increase towards older epochs \citep[e.g.][]{yamada05,chelouche07,planck16104P}. 
 In terms of visual extinction, the $A_{V}$ values attributed to the presence of ICD vary significantly in the range $0.004 < A_{V} < 0.5$, mainly due to the large measurements uncertainties \citep[e.g.][]{muller08}. However, despite these fluctuations in value, all studies tend to agree that there is only a small amount of dust in the ICM of clusters that can reach 1-3\% of the Galactic value \citep[e.g.][]{chelouche07,giard08,planck16104P,gjergo18,vogelsberger19}.\\
\noindent
In this letter, we present the analysis of the near-UV - $i$ $(\mathrm{NUV} -i)$ reddening of background galaxies in the Virgo direction and present the first detection of the diffuse dust component in the Virgo cluster over a radial scale of 0.4 virial radii. The Virgo system is still in the process of forming \citep[e.g.][]{conselice01,boselli14}, as witnessed by its irregular structure, which is characterised by a major sub-cluster A around the giant elliptical galaxy M87, and two smaller and less dense sub-clusters B and C around the cluster galaxies M49 and M86, respectively \citep{binggeli87,nulsen95}. Several optical studies have detailed the presence of an IC stellar population \citep{arnaboldi02,aguerri05,doherty09,durrell14,hartke17,longobardi13,longobardi15a,longobardi18a,longobardi18b,mihos17}, which is the result of pre-merger interactions and tidal stripping of cluster galaxies, and an FIR analysis has shown that gravitational interactions may also be responsible for an IC stream of dust in the ICM of sub-cluster C \citep{stickel03}.\\
\noindent 
Throughout the paper, we consider the cluster centred on M87, with a virial radius $r_{vir} = 1.55 $ Mpc \citep{mcLaughlin99}, and assume a distance for Virgo of 16.5 Mpc \citep{mei07,blakeslee09}, implying a physical scale of 80 pc arcsec$^{-1}$.

\section{Evidence for diffuse dust in the intra-cluster space of Virgo}
\label{sec2}
\subsection{Contamination from Galactic cirrus in the Virgo cluster area}
The GUViCS survey \citep[GALEX Ultraviolet Virgo Cluster Survey;][]{boselli11} presents GALEX UV observations of the Virgo cluster. It combines data from the All-sky Imaging Survey (AIS; $\sim$ 5\arcsec spatial resolution and single-exposure times of typically 100 s) and the Medium Imaging Survey (MIS; same spatial resolution, but with deeper exposure times of at least 1500 s). With this information, \citet{boissier15} produced FUV and NUV mosaics with 1\arcmin and 20\arcsec sized pixels that used the AIS and the MIS data, respectively, and showed that the FUV brightness resulting from the scattering of starlight by dust grains successfully traces the Milky Way (MW) dust column density. They then studied the variation of the average $(\mathrm{NUV}-i)$ colours of Sloan Digital Sky Survey (SDSS) background galaxies as a function of the Galactic reddening, $E(B-V)_G$, derived from either their FUV maps, the \citet{schlegel98}, or the \citet{planck14571A} maps, and showed that a correlation is present independently of the considered source of dust reddening. This result provides us with the opportunity of studying the distribution of the background galaxy colours after correction for Galactic contamination. To do so, we used a large sample of background sources from the \citet{voyer14} catalogue that provides us with NUV GUViCS and optical SDSS photometry, as well as redshift information. The data extend out to a maximum distance of $\sim 14$ deg from the centre of Virgo, but we restricted our sample to the objects lying within 6 deg ($1.1\times r_{vir}$), where the data are more homogeneous in terms of spatial coverage. Furthermore, we only considered sources with uncertainties on the magnitudes smaller than 0.1 mag, resulting in a final sample of $\sim 12000$ objects in the redshift range $0.02 < z < 0.8$. Then, we corrected the $(\mathrm{NUV}-i)$ colours for Galactic reddening through the formula    
\begin{equation}
({\rm NUV}-i) - ({\rm NUV}-i)_{c} = R({\rm NUV}-i) \times E(B-V)_G,
\label{reddening}
\end{equation} 
where $({\rm NUV}-i)$ and $({\rm NUV}-i)_{c}$ are the measured and the de-reddened galaxy colours, $E(B-V)_G$ is the Galactic reddening measured by \citet{boissier15} using different MW dust tracers, and ${\rm R}({\rm NUV}-i) = {\rm R}({\rm NUV})-{\rm R}(i)$ is the reddening in $({\rm NUV}-i)$ relative to $E(B-V)_G$ and equal to ${\rm R}({\rm NUV})$ = 8.64 and ${\rm R}(i)$ = 1.69, based on an MW extinction curve \citep{fitzpatrick07,schlafly11}. 
\begin{figure}
\centering
\includegraphics[width=9.cm]{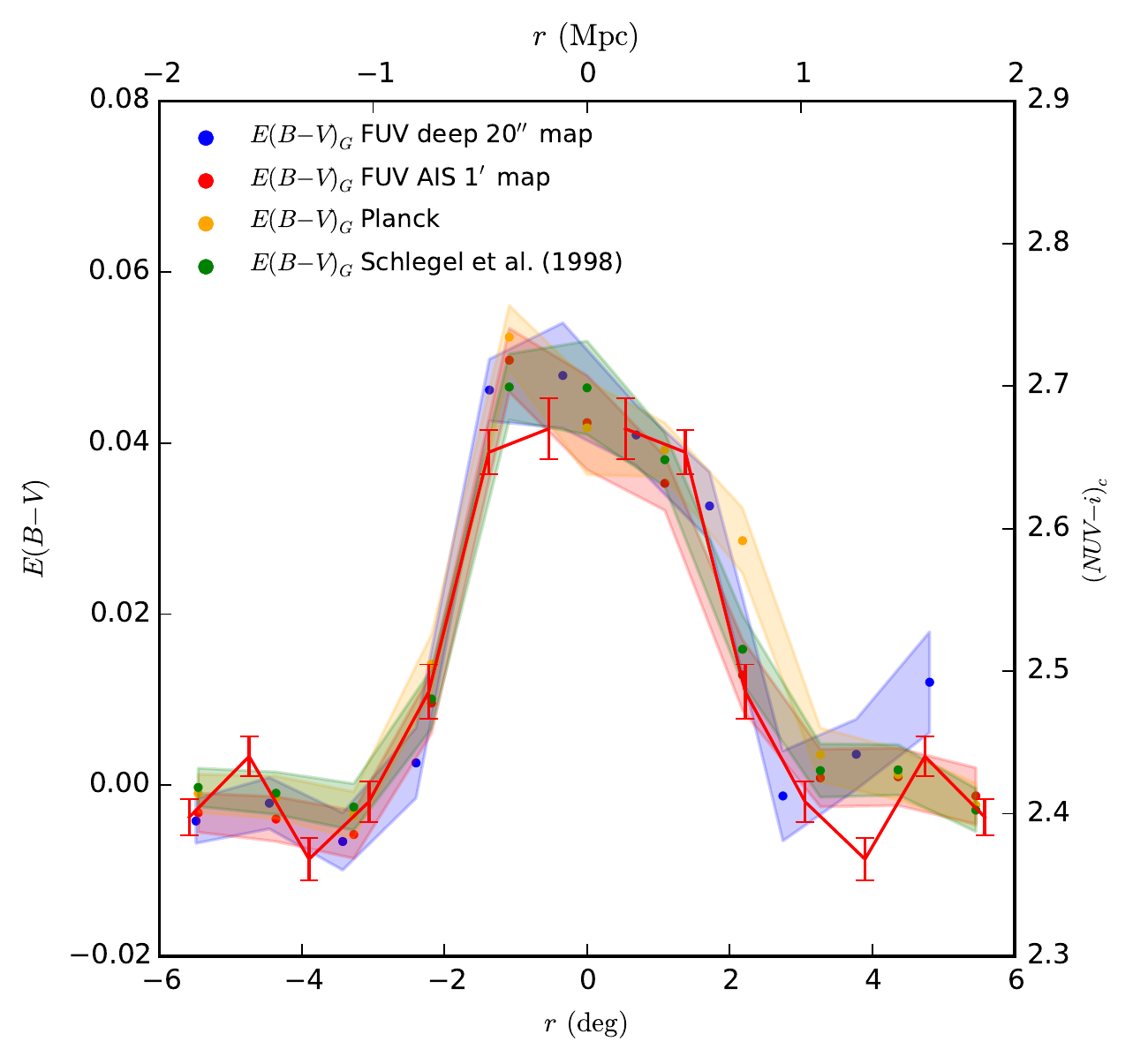}\\
\caption{\small{Radial profile of the $({\rm NUV}-i)_{c}$ colours and of the cluster $E(B-V)$ (filled dots) after the application of foreground extinction corrections, with their uncertainties shown as shaded areas. Positive and negative distances trace the northern and southern halves of the field. Different colours represent the profiles obtained by subtracting the Galactic extinction derived from the Schlegel (green), $Planck$ (yellow), FUV 1\arcmin (red), or 20\arcsec (blue) resolution maps. The continuous red line, with error bars, traces the radial profile obtained with no differentiation between the southern and northern halves of the field for the case where the Galactic extinction was derived from the FUV 1\arcmin map.}}
 \label{Colour_EBV_profile}
   \end{figure}

\subsection{Spatial variation of the intra-cluster dust in Virgo}
\label{Spataial_variation}
\begin{figure*}
\centering
\includegraphics[width=18.cm]{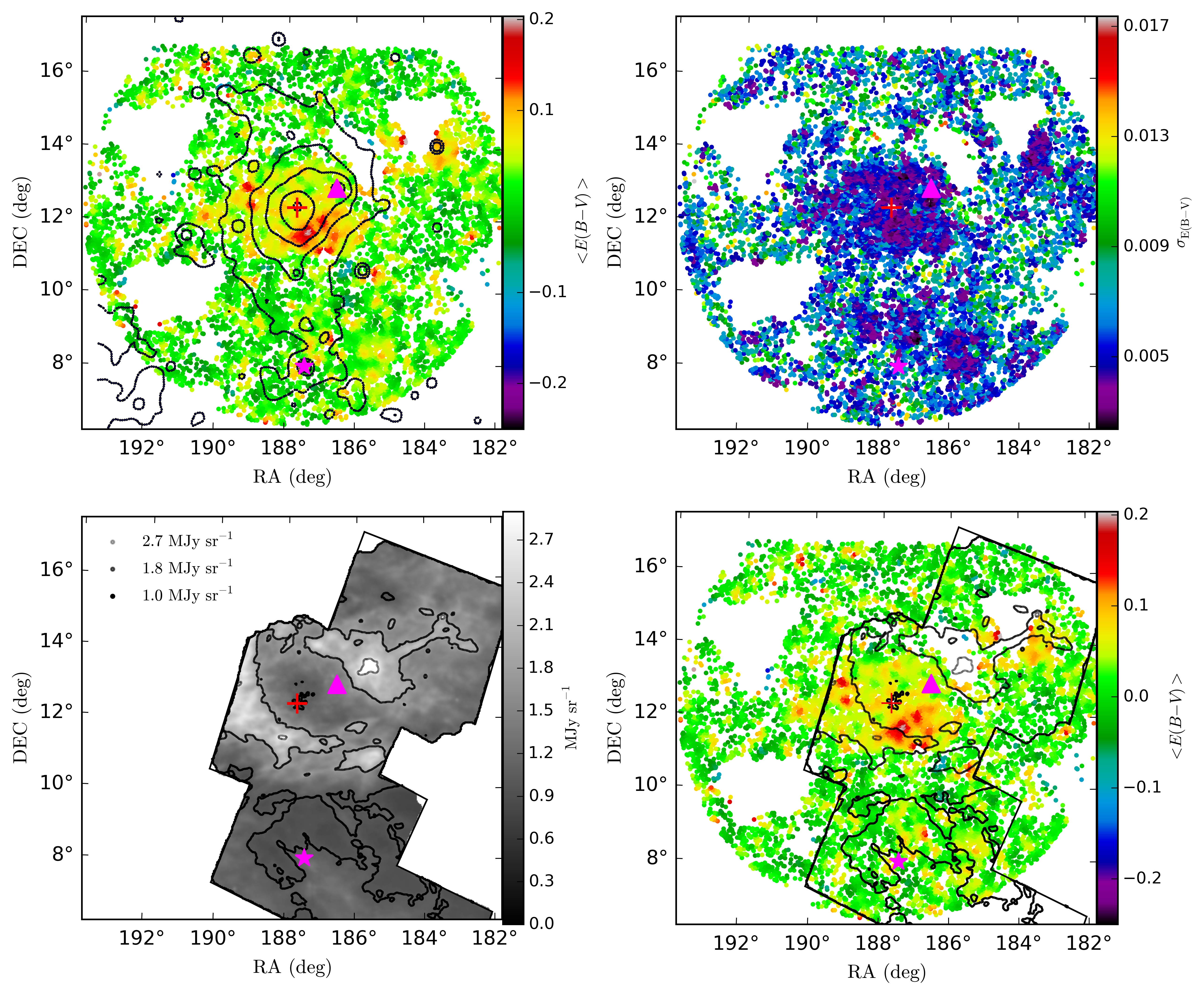}\\
\caption{\small{{\bf Top left panel:} Spatial distribution of the background galaxies, colour-coded according to the cluster $E(B-V)$ means. The peak measured south-west of M87 (red cross) is caused by background clusters. Black contours trace the cluster X-ray emission \citep{boehringer94}. {\bf Top right panel:} Errors in the reddening map computed by means of Monte Carlo simulations. {\bf Bottom left panel:} Model of the Galactic dust emission in the HeViCS fields in the SPIRE 250 $\mu m$ band \citep{bianchi17}. {\bf Right panel:}  Comparison between the Virgo dust reddening and different levels of Galactic contamination (see legend in the bottom left panel). In all panels, the red cross identifies the centre of the cluster, while the magenta star and triangle show the positions of M49 and M86, respectively.}}
 \label{Maps}
   \end{figure*}
We aim to detect the Virgo ICD by measuring its attenuation through the spatial variation in the de-reddened $(\mathrm{NUV}-i)_{c}$ colours of the background galaxies. The results are given in terms of the $(\mathrm{NUV}-i)_{c}$ colours as obtained by subtracting the Galactic extinction traced by the GALEX FUV AIS 1\arcmin map, but the same results are obtained when other Galactic extinction corrections are used (see Fig.\ref{Colour_EBV_profile} and Fig.\ref{radial_profiles}).\\
The cluster $E(B-V)$ values were derived through the relation 
\begin{equation}
 E(B-V) = \frac{({\rm NUV}-i)_{c} - ({\rm NUV}-i)_{0}}{R({\rm NUV}-i)_{c}},
\label{cluster_extinction}
\end{equation}
with $R({\rm NUV}-i)_{c} = 7.8$, based on the extinction curve of the Large Magellanic Cloud (LMC \citep{gordon03}. (Progenitors of the Virgo intra-cluster component are LMC-like systems.) $({\rm NUV}-i)_{0}$ represents the intrinsic colour of the objects if no obscuration by the cluster dust was present. It was fixed to the average of the $(\mathrm{NUV}-i)_{c}$ values outside 4 deg from the cluster centre where its profile flattens, as shown in Fig.\ref{Colour_EBV_profile}. (We verified that there is no variation in the background value of the $(\mathrm{NUV}-i)_{c}$ colours out to 14 deg, $\sim 2.6 \times r_{vir}$.) Fig.\ref{Colour_EBV_profile} also shows the variation with radius of the mean cluster $E(B-V)$ values after the application of different corrections for foreground Galactic extinction. Because we expect the initial estimate of the mean to be influenced by the presence of background clusters that are dominated by red quiescent galaxies that can mimic higher $E(B-V)$ values \citep{boselli06}, we computed an iteratively sigma-clipped mean by applying a 2$\sigma$ limit. The relation is plotted for positive and negative $r$ to trace the northern and southern halves with respect to the centre of Virgo (dots with shaded areas), and in the case of no differentiation between the two sides of the cluster (red continuous line with error bar).
The resulting profile shows a component that for $r \le$ 1.5 deg ($\sim$ 0.4 Mpc or 0.3 $r_{vir}$) is constant, within the uncertainties, and is characterised by an average reddening $E(B-V)\sim 0.042\pm 0.004$. This result is in agreement with observational and theoretical studies of dust distribution in galaxy clusters, which attribute the flattening to sputtering phenomena that more efficiently destroy the dust closer to the centre \citep[e.g.][]{bai07,gutierrez17,vogelsberger19}. Beyond 1.5 deg, the extinction starts to decrease and reaches levels of $E(B-V)\sim 0.023\pm 0.004$ at $\sim$ 2.0 deg ($\sim$ 0.6 Mpc or 0.4 $r_{vir}$), and it is consistent with zero for larger distances.\\
\noindent
Next, we built a two-dimensional map of the cluster reddening performing a Gaussian smoothing of the $E(B-V)$ values at the position of every $\sim$ 12000 background galaxies, also weighted by the uncertainty on the measurement. To suppress small-scale variations, the smoothing length was set to $h=6\arcmin$ ($\sim$ 30 kpc). However, the final results are not strongly sensitive to the bandwidth size. Errors in the derived smoothed map were computed by means of Monte Carlo simulations by generating 100 different data sets of mock $E(B-V)$ values with the same positions on the sky as for the real sample of background galaxies. Simulated reddenings were calculated from the measured two-dimensional reddening map by adding a random value that resembled the associated measurement error. Because the same smoothing procedure was applied to the synthetic data, the statistics of these simulated $E(B-V)$ fields give us the error associated with the smoothed extinction values at the source positions in our field. 
The results are shown in Fig.\ref{Maps}, where we plot the position on the sky of the background galaxies, colour-coded based on their mean $E(B-V)$ and their errors (top panels). In agreement with what is shown in Fig.\ref{Colour_EBV_profile}, these maps suggest that reddening by dust is statistically significant within 2 deg from the centre of Virgo and that it is enhanced within a circular region around the dynamical centre of the cluster (sub-cluster A), as identified by the X-ray gas emission (black contours). Interestingly, no dust attenuation is measured in correspondence of the ellipticals M49 (magenta star) and M86 (magenta triangle), centres of Virgo sub-clusters B and C, respectively. The high values visible in a small south-west region are due to the presence of background clusters whose systems are intrinsically redder in (NUV-$i$) colours and hence should not be considered as tracing reddening by dust. This was verified by transforming our set of RA, DEC, and redshift into points in the 3D Cartesian space and then examining statistically significant over-densities in three-dimensional space. Finally, to verify that the signal we measured was not due to residual contamination by Galactic cirrus, we compared it with the MW dust emission in the SPIRE 250 $\mu$m band \citep{bianchi17} in the region where they overlap. The comparison is shown in Fig.\ref{Maps} (bottom panels), where it is evident that the spatial distribution of the Galactic FIR emission differs from the distribution of the cluster dust extinction.


\section{Physical properties of the Virgo dust}
\label{sec3}
\paragraph{{\bf Infrared flux density and dust temperature}}
The dust IR flux density is related to the optical reddening as ${I_{\nu}= \tau_{\nu_{0}}\left(\frac{\nu}{\nu_{0}}\right)^{\beta}B_{\nu}(T_{d})}$, where $B_{\nu}(T_{d})$ is the Planck function for dust at temperature $T_{\mathrm{d}}$, $\beta = 1.5$ is the adopted dust emissivity index, and $\tau_{\nu_{0}} = 0.67 \times 10^{-4} \times E(B-V)$ is the dust optical depth at the reference frequency ${\nu_{0}} = 353$ GHz \citep{planck14571A}. By analysing $Herschel$ data, \citet{bianchi17} found as upper limit on the Virgo ICD emission at 250 $\mu m$, $I_{250\mu m} \sim $ 0.1 $\mathrm{MJy\, sr^{-1}}$. Based on our $E(B-V)$ estimates and assuming the Galactic values for $\beta$ and $\tau_{\nu_{0}}$, this result implies an upper limit for the temperature for the Virgo diffuse dust of $T_{\mathrm{d}} \sim 10\, \mathrm{K}$. The dust temperature distribution is expected to peak around these values in Virgo-like environments \citep{popescu00}, and the detection of cold grains is not limited to this study. \citet{fogarty19} found dust at $T_{d}\sim10$ K for clumps of material thtat extended several kiloparsecs away from the optical peak of the galaxy, and the authors concluded that these grains must either be shielded from sputtering or lie in regions of cooler ICM plasma. This result should be taken with some caution, however, because our estimated results are derived under the assumptions that the Virgo ICD exhibits similar emissivity as the MW. Similar values of flux densities could be obtained for dust grains with lower emissivities at higher temperatures.\\
\noindent
Under the assumption of $T_{\mathrm{d}} \sim 10$ K, we plot in Fig.\ref{radial_profiles} the radial variation of the IR flux density at 250 $\mu m$ in the radial range where we measure positive $E(B-V)$ values. Within 1.5 deg, the profile remains nearly constant to a value of $0.14\pm0.01\, \mathrm{MJy\, sr^{-1}}$, and it decreases to $0.08\pm0.01\, \mathrm{MJy\, sr^{-1}}$ at $\sim$ 2 deg. The general agreement with the surface brightness (SB) profiles of the X-ray gas \citep{simionescu17} and of the optical ICL is good \citep{longobardi18b} (black and purple lines, respectively). However, closer to the centre, dust may be more efficiently destroyed \citep[see Sect.\ref{Spataial_variation}, but also][]{vogelsberger19}, resulting in a deficit with respect to the stellar and hot gaseous components. The comparison with the SB profile of the Virgo ICL can help us to understand the origin of the diffuse ICD.  Previous works that used PNs and GCs to dynamically trace the ICL have shown that the IC population in Virgo is the accreted component of the cluster from low- and intermediate-mass star-forming and dwarf-ellipticals galaxies. It therefore exhibits a centrally concentrated profile as a consequence of the dynamical friction that drags the satellite galaxies towards the dynamical centre, and baryons are stripped and orbit the cluster as the IC component.  In parallel, studies of the dust content in cluster members have shown that systems approaching regions of high density are found to be redder and dust deficient than the population of galaxies in the field \citep{gavazzi10,cortese12}. This scenario is even more dramatic for low-mass-galaxies \citep{boselli14}. If then the Virgo ICL is built up predominantly by tidal stripping of low-mass objects, we do expect the presence of dust in the IC space. Removed from the cluster galaxies and transported to the IC component by the same environmental processes that remove their stellar content and/or by additional ram pressure phenomena \citep{stickel03,stickel05,cortese10,cortese12}, it would have similar spatial distributions, as our analysis shows.
\vspace{-0.5cm}
\paragraph{{\bf Visual extinction, mass, and dust-to-gas ratio }}
From the $E(B-V)$ values presented in Sect.\ref{sec2} and assuming an LMC extinction factor $R_{V} = 3.4$ \citep{gordon03}, we derive an average visual extinction for the Virgo ICD, $A_{V} = 0.14\pm0.01$ within 1.5 deg from the centre that decreases to $A_{V} = 0.08\pm0.01$ at 2 deg. The very low $A_{V}$ values that we find to characterise the Virgo ICD make it clear that the detection of such a component was only possible through the availability of NUV data, which are very sensitive to extinction by dust.
Within the uncertainties of our measurements, Fig.\ref{Colour_EBV_profile} also shows that the ICD is uniformly distributed for $ r \le 1.5\, \mathrm{deg}$, allowing us to relate the average visual extinction to the total mass of the ICD, $M_{d}$ \citep{muller08}. Under the assumption that the Virgo dust grains are standard silicate grains, our result yields a total mass $M_{d} = 2.5\pm 0.2 \times10^{9}M_{\odot}$. 
We next computed the $M_{d}/M_{g}$ profile of the dust-to-gas ratio. The estimate of the cluster gas mass in radial bins is obtained through the formula $M_{gas}= 4\pi\int_{r_{i}}^{r_{i+1}}{\mu_{e} m_{u} n_{e}(r)r^{2} dr}$ \citep{ettori13}, where $\mu_{e} = 1.155$, $m_{u} = 1.6610^{-24}\, \rm{g}$ is the atomic mass unit, and $n_e(r)$ is the Virgo electron number density \citep{simionescu17}. The results, plotted in Fig.\ref{radial_profiles}, show no significant radial gradient, characterised by an average value of $M_{d}/M_{g} = 3.0\pm 0.3 \times 10^{-4}$. \citet{montier04} have shown that when dust grains are in the ICM, they can dominate the cooling mechanisms if the gas temperature is higher than $T_{g} \ge 10^{7} $ K and $M_{d}/M_{g} > 2\times 10^{-5}$. Thus, the ICD in Virgo acts as a gas coolant and contributes to a change in the statistical properties of the Virgo ICM \citep[e.g.][]{montier04,pointecouteau09,vogelsberger19}.
\vspace{-0.5cm}
\paragraph{{\bf Background obscuration}}
The identification of an ICD component in Virgo has consequences for the count of background sources in the Virgo direction. For a uniform dust optical depth $\tau_{V} = 0.92 A_V$, the lost fraction of background galaxies can be approximated as $f_{miss} = 1- e^{\alpha \tau_{V}}$ \citep{masci98}, with $\alpha$ the slope of the optical luminosity function (LF) of the galaxies. When we assume $\alpha \sim -1.3$ for galaxies with $z$ in the range of our sample \citep{ilbert05}, this yields an $\sim 15\%$ deficit within 1.5 deg of Virgo. Larger deficits are expected for sources at higher $z$ given the evidence of steeper LFs for systems at younger epochs.


\section{Summary and conclusions}
\label{conclusions}
In this letter, we have presented the first finding of diffuse IC dust in the ICM of the Virgo cluster over a radial scale of 0.4 virial radii. The ICD was detected by measuring (NUV-$i$) colour excess for a sample of $\sim 12000$ background galaxies that lie within 6 deg (1.1 $r_{vir}$) from the cluster centre and whose magnitudes in the GUViCS NUV- and SDSS $i$-bands were measured with an uncertainty $\sigma< 0.1$ mag \citep{voyer14}. The contamination by Galactic cirrus was subtracted by de-reddening the measured colours of the correlation found to relate the ($\mathrm{NUV}-i$) colours and the Galactic dust reddening \citep{boissier15}. 
By assuming an LMC extinction curve, we find that the estimated colour excess translates into $E(B-V)$ values such that closer to the centre and within a 1.5 deg radius (0.3 $r_{vir}$), the optical reddening of the ICD is $E(B-V)\sim0.042 \pm 0.004$ mag. More specifically, our results show the following:
\begin{figure}
\centering
\includegraphics[width=9.2cm]{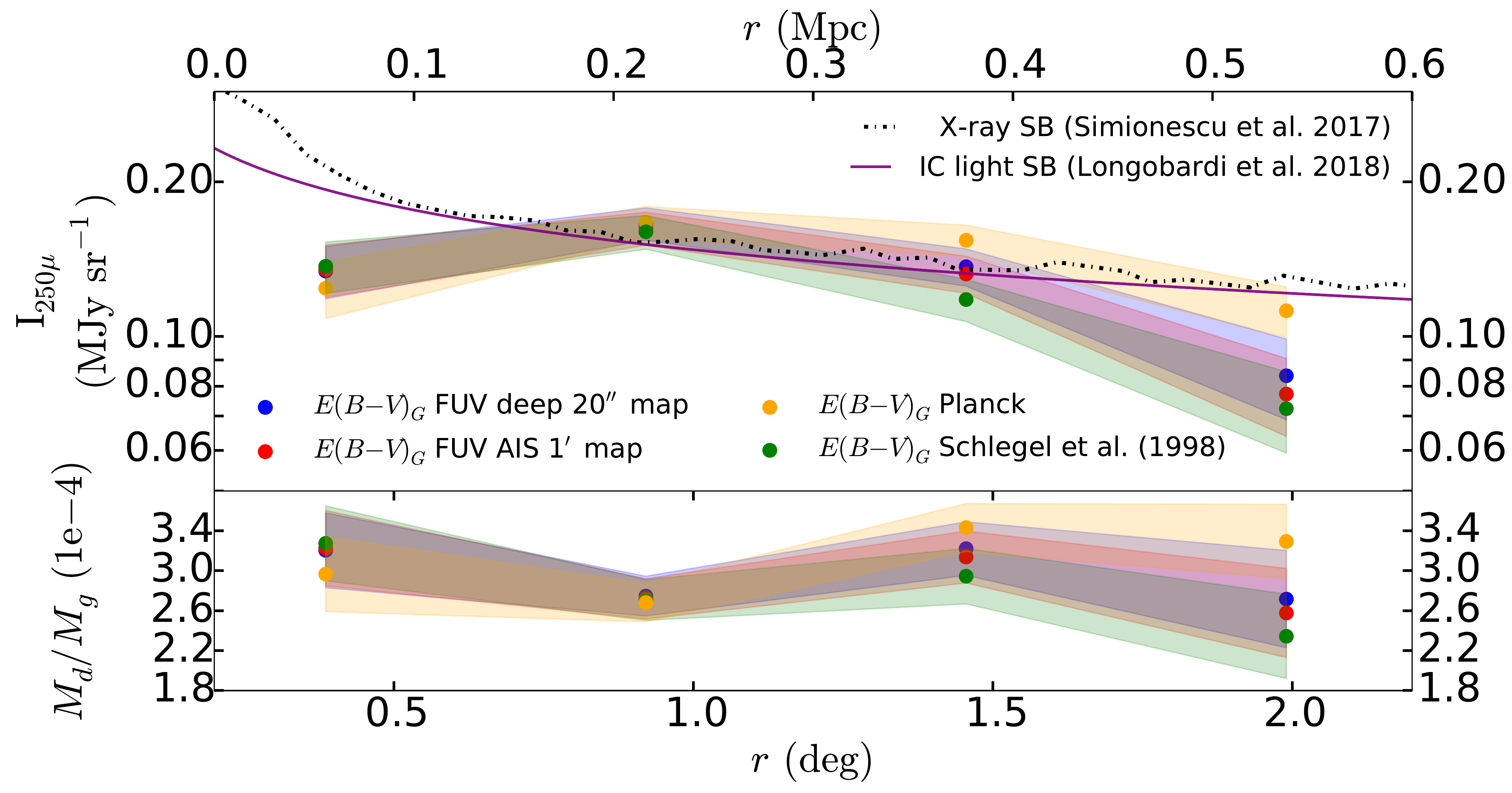}\\

\caption{\small{Radial variation of the Virgo ICD properties (dots).{\bf Top panel}: Expected dust intensities at 250 $\mu m$ as obtained from our $E(B-V)$ measurements. The dash-dotted and full purple lines denote the SB profiles of the X-ray gas and optical IC light components, respectively. They have been scaled for a direct comparison with the $I_{250_{\mu m}}$ profile. {\bf Bottom panel}: Dust-to-gas mass ratios. The different colours identify values derived using different Galactic extinction corrections. Shaded areas trace the uncertainties on the mean values that include the standard deviation and the intrinsic measurement errors.}}
 \label{radial_profiles}
   \end{figure}
\vspace{-0.1cm}
\begin{itemize}
\item Virgo has a diffuse ICD in a central region with a radius of $\sim$2.0 deg that increases towards the cluster centre as measured for the ICL component, although the variation with radius is shallow within $\sim$ 1.5 deg (0.3 $r_{vir}$). The similar spatial distributions of the Virgo ICD and ICL suggest that diffuse dust in Virgo is transported through similar processes as those that originate the IC stars.

\item  The reddening values we find to characterise the diffuse dust in Virgo imply variations in the physical properties of the Virgo ICD with respect to the Galactic values either in terms of temperature (colder for the ICD) or in terms of emissivity (higher for the MW). In both cases, our findings emphasise the need for multi-band data observations to further constrain the physical properties of dust grains outside the main body of galaxies.

\item Within 1.5 deg (0.3 $r_{vir}$), the Virgo ICD is characterised by an average extinction, $A_{V} = 0.14\pm0.01$. This results in a total dust mass $M_{d} = 2.5\pm0.2\times10^{9}M_{\odot}$ that is widespread in the Virgo ICM with a dust-to-gas mass ratio $M_{d}/M_{g} = 3.0\pm 0.3 \times 10^{-4}$. This means that it is deficient by a factor of 100 relative to the Galactic ISM value and implies that the ICD in Virgo constitutes an additional cooling agent of the gas. Thus, the ICD plays an important role in the ICM physics of the Virgo cluster.

\item  The presence of diffuse ICD obscures background sources and may bias optical studies in the direction of Virgo.
\end{itemize}
\noindent
This study is important for missions such as Euclid, Athena, and SPICA, which are powerful tools for constraining dust-gas interactions in the IC space of clusters.

\begin{acknowledgements}
We thank the anonymous referee for helpful comments. AL and AN acknowledge support from the French Centre National d'Etudes Spatiales (CNES).
\end{acknowledgements}

\bibliographystyle{aa}
\bibliography{Biblio}

\end{document}